\documentstyle[aps]{revtex}
\begin{document}
\title{Unitary Random-Matrix Ensemble with Governable Level Confinement}
\author{\bf V. Freilikher, E. Kanzieper}
\address{The Jack and Pearl Resnick Institute of Advanced Technology,\\
Department of Physics, Bar-Ilan University, Ramat-Gan 52900, Israel}
\author{\bf I. Yurkevich$^{\#}$}
\address{The Jack and Pearl Resnick Institute of Advanced Technology,\\
Department of Physics, Bar-Ilan University, Ramat-Gan 52900, Israel\\
and International Centre for Theoretical Physics, Trieste 34100, Italy}
\date{October 2, 1995}
\maketitle

\begin{abstract}
A family of unitary $\alpha $-Ensembles of random matrices with governable
confinement potential $V\left( x\right) \sim \left| x\right| ^\alpha $ is
studied employing exact results of the theory of non-classical orthogonal
polynomials. The density of levels, two-point kernel, locally rescaled
two-level cluster function and smoothed connected correlations between the
density of eigenvalues are calculated for strong $\left( \alpha >1\right) $
and border $\left( \alpha =1\right) $ level confinement. It is shown that
the density of states is a smooth function for $\alpha >1$, and has a
well-pronounced peak at the band center for $\alpha \leq 1$. The case of
border level confinement associated with transition point $\alpha =1$ is
reduced to the exactly solvable Pollaczek random-matrix ensemble. Unlike the
density of states, all the two-point correlators remain (after proper
rescaling) to be universal down to and including $\alpha =1$.

PACS number(s): 05.45.+b, 05.40.+j
\end{abstract}

%\twocolumn

\section{\bf Introduction}

Random matrix theory\cite{Mehta} pretends to description of a great variety
of physical systems from complex nuclei and classically chaotic systems to
electron transport in disordered conductors. The latters are known to
exhibit crossover from metallic phase with Wigner-Dyson level statistics to
insulating one with Poisson level statistics as disorder increases, passing
through the Anderson transition where a third universal statistics should
exist \cite{Shklovskii and Kravtsov}.

There are two ways to describe such a crossover within the random matrix
theory. The first one deals with random matrix ensembles which do not hold
their invariance under orthogonal, unitary or symplectic transformations
from the beginning \cite{Shapiro} due to the presence of a symmetry-breaking
term in the joint probability density function $P\left[ {\bf H}\right] $ of
the $N\times N$ elements of the random matrix ${\bf H}$. Such random matrix
ensembles with primordially broken symmetry demonstrate a deviation of the
level statistics from the Wigner-Dyson type to the Poissonian one.

The second approach starts with invariant ensembles of random matrices that
implies the logarithmic repulsion between the levels confined by a
parameter-dependent potential. In this case \cite{Mehta} a rigorous
treatment in the terms of orthogonal polynomials turns out to be successful
if those are known for the chosen confinement potential. For instance, in
the model based on the non-classical $q$-polynomials \cite{Muttalib} the
relevant parameter, entering the confinement potential, is associated with
the strength of disorder. The strong confinement was found to be relevant to
the weak disorder (metallic regime), while soft confinement potential was
characteristic for strong disorder (insulating regime).

Correspondence between these two approaches has been traced in \cite
{Canalli-Kravtsov's reprint}, where it was shown that in random-matrix
ensemble with quadratic logarithmic confinement potential the spontaneous
breaking of underlying symmetry of $P\left[ {\bf H}\right] $ occurs. The
symmetry breaking manifests itself in the loosing of the translational
invariance of the two-level cluster function $Y_2\left( s,s^{\prime }\right)
$ and leads to the Poisson-like level statistics.

Another family of unitary ensembles of random matrices with governable
confinement potential was proposed in \cite{Manning}, where so-called $%
\alpha $-Ensemble was introduced. Symmetric $\alpha $-Ensemble was
considered in \cite{Canalli-Wallin-Kravtsov}. This ensemble provides an
excellent basis for rigorous treatment\cite{FKY} and enables to explore how
soft the confinement potential must be to cause deviations from the
Wigner-Dyson statistics.

Let us consider a physical system with broken time-reversal symmetry
described by an $N\times N$ random matrix ${\bf H}$ with eigenvalues $%
\left\{ x_n\right\} ,n=1,...,N$. The joint probability density function $%
P\left( \left\{ x\right\} \right) $ can be written in the form \cite{Mehta}
\begin{equation}
P\left( \left\{ x\right\} \right) =Z^{-1}\exp \left( -\beta \left(
\sum_iV\left( x_i\right) -\sum_{i<j}\ln \left| x_i-x_j\right| \right)
\right) ,\text{ }\beta =2,  \label{eq.7}
\end{equation}
which implies a pairwise logarithmic repulsion between the levels confined
by the potential $V\left( x\right) $; $Z$ is a partition function.

Symmetric $\alpha $-Ensemble is characterized by parameter-dependent
potential
\begin{equation}
V_\alpha \left( x\right) =\frac 12\left| x\right| ^\alpha  \label{eq.001}
\end{equation}
supported on the whole real axis $x\in \left] -\infty ,+\infty \right[ $.

The Monte Carlo simulations\cite{Canalli-Wallin-Kravtsov}, based on the
mapping of initial random-matrix problem onto the problem of interacting
particles in confined $1D$ plasma, showed that in such $\alpha $-Ensemble
the deviations from the Wigner-Dyson statistics may also occur, and they
take place near the center of the spectrum for $0<\alpha <1$. The value $%
\alpha =1$ was associated with the sharp transition point corresponding to
the crossover from strong to soft level confinement. Just at this point, $%
\alpha =1$, the mean-field approximation widely used in the random-matrix
theory fails, giving rise to the singularity of the level density near the
spectrum origin.

In this paper we present the analytical treatment of $\alpha $-Ensemble. In
Section II $\alpha $-Ensemble with strong confinement potential $\left(
\alpha >1\right) $ is studied employing exact results of the theory of
non-classical orthogonal polynomials. In Section III we describe the general
properties of symmetric $\alpha $-Ensemble revealing the fact that $\alpha =1
$ is a special point associated with peak formation in the density of
states. This special case of {\it border} level confinement is investigated
analytically in the Section IV in the framework of the theory of orthogonal
polynomials, exploiting the properties of the symmetric Pollaczek
polynomials. In Sections III and IV we find the density $\nu _N\left(
x\right) $ of levels, two-point kernel $K_N\left( x,y\right) $, locally
rescaled two-level cluster function $Y_2\left( s,s^{\prime }\right) $ and
smoothed connected correlations $\left[ \nu _N\left( x,y\right) \right]
_{con}$ between the density of eigenvalues. It will be shown that for strong
and border level confinement the rescaled local two-level cluster functions
as well as the smoothed connected correlations follow the universal forms
which are typical of random-matrix ensembles with the Wigner-Dyson level
statistics \cite{Brezin-Zee},\cite{Beenakker},\cite{Weidenmuller}. Section V
contains conclusions.

\section{Symmetric $\alpha $-Ensemble: case of strong level confinement $%
\left( \alpha >1\right) $}

The rigorous treatment of $\alpha $-Ensemble with strong level confinement
has been made possible by the recent development of the theory of
non-classical orthogonal polynomials. The relevant polynomials $P_n^{\left(
\alpha \right) }\left( x\right) $ are known as those orthogonal with respect
to the Freud weights \cite{Nevai}, \cite{Lubinsky}.

All the $n$-point correlation functions can be expressed through the
two-point kernel \cite{Mehta}
\begin{equation}
K_N^{\left( \alpha \right) }\left( x,y\right) =e^{-V_\alpha \left( x\right)
-V_\alpha \left( y\right) }\frac{k_{N-1}}{k_N}\frac{P_{N-1}^{\left( \alpha
\right) }\left( x\right) P_N^{\left( \alpha \right) }\left( y\right)
-P_{N-1}^{\left( \alpha \right) }\left( y\right) P_N^{\left( \alpha \right)
}\left( x\right) }{y-x},  \label{eq.7L}
\end{equation}
where $k_N$ is a leading coefficient of the orthonormal polynomial $%
P_N^{\left( \alpha \right) }\left( x\right) $. The pointwise asymptotics of $%
P_N^{\left( \alpha \right) }\left( x\right) $ for large $N$ have the form
\cite{Rahmanov}
\begin{equation}
P_N^{\left( \alpha \right) }\left( x\right) =\left( \frac 2{\pi D\left(
N,\alpha \right) }\right) ^{1/2}\frac{\exp \left( V_\alpha \left( x\right)
\right) }{\left[ 1-\left( x/D\left( N,\alpha \right) \right) ^2\right] ^{1/4}%
}\cos \left( \Phi _N^{\left( \alpha \right) }\left( x\right) \right) ,
\label{eq.8L}
\end{equation}
with
\begin{equation}
\Phi _N^{\left( \alpha \right) }\left( x\right) =\pi N\int_{x/D\left(
N,\alpha \right) }^1\omega _\alpha \left( z\right) dz+\frac 12\arccos \left(
\frac x{D\left( N,\alpha \right) }\right) -\frac \pi 4.  \label{eq.9L}
\end{equation}
Here
\begin{equation}
\omega _\alpha \left( z\right) =\frac \alpha \pi \left| z\right| ^{\alpha
-1}\int_{\left| z\right| }^1\frac{d\eta }{\eta ^\alpha \sqrt{1-\eta ^2}}
\label{eq.10L}
\end{equation}
is the Nevai-Ullman density,
\begin{equation}
D\left( N,\alpha \right) =N^{1/\alpha }D\left( \alpha \right) ,\text{ }%
D\left( \alpha \right) =\left( \frac{2^{\alpha -1}\Gamma ^2\left( \frac
\alpha 2\right) }{\Gamma \left( \alpha \right) }\right) ^{1/\alpha },
\label{eq.11L}
\end{equation}
and
\begin{equation}
k_{N-1}/k_N=\frac 12D\left( N,\alpha \right) .  \label{eq.12L}
\end{equation}

The straightforward calculations based on Eqs. (\ref{eq.7L}) - (\ref{eq.12L}%
) lead to the following expression:
\begin{eqnarray}
K_N^{\left( \alpha \right) }\left( x,y\right) &=&\frac 1{\pi \left(
y-x\right) }\left[ \left( 1-\left( \frac x{D\left( N,\alpha \right) }\right)
^2\right) \left( 1-\left( \frac y{D\left( N,\alpha \right) }\right)
^2\right) \right] ^{-1/4}  \label{eq.13L} \\
&&\times \left[ \cos \left( \Phi _{N-1}^{\left( \alpha \right) }\left(
x\right) \right) \cos \left( \Phi _N^{\left( \alpha \right) }\left( y\right)
\right) -\cos \left( \Phi _N^{\left( \alpha \right) }\left( x\right) \right)
\cos \left( \Phi _{N-1}^{\left( \alpha \right) }\left( y\right) \right)
\right] .  \nonumber
\end{eqnarray}

Despite of this function is rather complicated, it can be rewritten in the
universal form if one supposes that $\left| x-y\right| $ is much smaller
than the scale $\varsigma $ of characteristic changes of the mean level
density, $\varsigma \sim \nu _N^{\left( \alpha \right) }\left( d\nu
_N^{\left( \alpha \right) }/dx\right) ^{-1}\sim D\left( N,\alpha \right) $.
Assuming also that both $x$ and $y$ stay away from the band edge $D\left(
N,\alpha \right) $ of the spectrum and making use of the asymptotic identity
$\Phi _{N-1}^{\left( \alpha \right) }\left( x\right) =\Phi _N^{\left( \alpha
\right) }\left( x\right) -\arccos \left( x/D\left( N,\alpha \right) \right) $%
, we obtain in the leading order in $1/N$
\begin{equation}
K_N^{\left( \alpha \right) }\left( x,y\right) =\frac{\sin \left( \pi
\overline{\nu }_N^{\left( \alpha \right) }\left( x-y\right) \right) }{\pi
\left( x-y\right) },  \label{eq.14L}
\end{equation}
where
\begin{equation}
\overline{\nu }_N^{\left( \alpha \right) }=\frac N{D\left( N,\alpha \right) }%
\omega _\alpha \left( \frac{x+y}{2D\left( N,\alpha \right) }\right)
\label{eq.15L}
\end{equation}
is the local density of levels. Note, that Eq. (\ref{eq.14L}) is valid on
the scale $\left| x-y\right| $ which is much larger than the mean
level-spacing $\left( \overline{\nu }_N^{\left( \alpha \right) }\right)
^{-1}\sim N^{1/\alpha -1}$.

Correspondingly, in the large-$N$ limit the two-level cluster function
\begin{equation}
Y_2\left( s,s^{\prime }\right) =\left( \frac{K_N^{\left( \alpha \right)
2}\left( x,y\right) }{\nu _N^{\left( \alpha \right) }\left( x\right) \nu
_N^{\left( \alpha \right) }\left( y\right) }\right) _{%
%TCIMACRO{\QATOPD. . {x=x\left( s\right) }{y=y\left( s^{\prime }\right) } }
%BeginExpansion
{x=x\left( s\right)  \atopwithdelims.. y=y\left( s^{\prime }\right) }
%EndExpansion
}\text{,}  \label{eq.17L}
\end{equation}
being rewritten in the terms of eigenvalues measured in the local level
spacing $s=x\overline{\nu }_N^{\left( \alpha \right) }$ and $s^{\prime }=y%
\overline{\nu }_N^{\left( \alpha \right) }$, locally{\it \ }follows the
universal form
\begin{equation}
Y_2\left( s,s^{\prime }\right) =\frac{\sin ^2\left[ \pi \left( s-s^{\prime
}\right) \right] }{\left[ \pi \left( s-s^{\prime }\right) \right] ^2}
\label{eq.18L}
\end{equation}
{\it irrespective} to the value $\alpha >1$.

As is well-known \cite{Mehta}, the level-spacing distribution function $%
P\left( \Delta \right) $ can be expressed through the eigenvalues $\left\{
\lambda \left( \Delta \right) \right\} $ of the Fredholm integral equation
where $\sqrt{Y_2}=\sin \left[ \pi \left( s-s^{\prime }\right) \right] /\pi
\left( s-s^{\prime }\right) $ stands for the kernel
\begin{equation}
\int_{-\Delta /2}^{\Delta /2}ds^{\prime }\sqrt{Y_2\left( s,s^{\prime
}\right) }f\left( s^{\prime }\right) =\lambda \left( \Delta \right) f\left(
s\right) ,  \label{eq.31L}
\end{equation}
\begin{equation}
P\left( \Delta \right) =\frac{d^2}{d\Delta ^2}\prod_j\left( 1-\lambda
_j\left( \Delta \right) \right) .  \label{eq.32L}
\end{equation}
Since $Y_2\left( s,s^{\prime }\right) $ for $\alpha $-Ensemble coincides
with that for Gaussian Ensemble, it inevitably leads to universal
Wigner-Dyson statistics.

The connected correlations between the density of eigenvalues for $x\neq y$
are given by\cite{Brezin-Zee}
\begin{equation}
\left[ \nu _N^{\left( \alpha \right) }\left( x,y\right) \right]
_{con}=-K_N^{\left( \alpha \right) 2}\left( x,y\right)  \label{eq.62aaL}
\end{equation}
and known to oscillate rapidly on the scale of the band width $D\left(
N,\alpha \right) .$ The smoothed correlation function which is useful for
the calculation of integral characteristic of spectrum can easily be
determined. Bearing in mind Eq. (\ref{eq.13L}) we obtain after averaging
over rapid oscillations
\begin{equation}
\left[ \nu _N^{\left( \alpha \right) }\left( x,y\right) \right] _{con}=-%
\frac 1{2\pi ^2\left( x-y\right) ^2}\frac{D\left( N,\alpha \right) ^2-xy}{%
\sqrt{\left( D\left( N,\alpha \right) ^2-x^2\right) \left( D\left( N,\alpha
\right) ^2-y^2\right) }}\text{, }x\neq y.  \label{eq.62abL}
\end{equation}
Equation (\ref{eq.62abL}) proves that smoothed correlations in $\alpha $%
-Ensemble with strong level confinement follow universal form\cite
{Brezin-Zee},\cite{Beenakker}.

Note that along with the proof of universality of eigenvalue correlations we
have obtained an {\it exact} formula for the density of levels in $\alpha $%
-Ensemble. Thus, Eqs. (\ref{eq.10L}) and (\ref{eq.15L}) yield following
expression for $\alpha >1$:
\begin{equation}
\nu _N^{\left( \alpha \right) }\left( x\right) =\frac{\alpha \Gamma \left(
\alpha \right) }{2^{\alpha -1}\pi \Gamma ^2\left( \frac \alpha 2\right) }%
\left| x\right| ^{\alpha -1}\int_{\left| z\right| }^1\frac{d\eta }{\eta
^\alpha \sqrt{1-\eta ^2}},  \label{eq.16L}
\end{equation}
that can be rewritten in two equivalent ways by means of Hypergeometric
functions,
\begin{equation}
\nu _N^{\left( \alpha \right) }\left( x\right) =\frac \alpha \pi \frac{%
\Gamma \left( \frac{1+\alpha }2\right) }{\sqrt{\pi }\Gamma \left( \frac
\alpha 2\right) }\left| x\right| ^{\alpha -1}\sqrt{1-z^2}\,_2{\bf F}_1\left(
\frac 12,\frac{1+\alpha }2;\frac 32;1-z^2\right) ,  \label{eq.16aL}
\end{equation}
or
\begin{equation}
\nu _N^{\left( \alpha \right) }\left( x\right) =\frac{\alpha N^{1-1/\alpha }%
}{\pi D\left( \alpha \right) }\sqrt{1-z^2}\,_2{\bf F}_1\left( 1,1-\frac
\alpha 2;\frac 32;1-z^2\right) ,  \label{eq.16bL}
\end{equation}
where $z=x/D\left( N,\alpha \right) $.

In the particular case $\alpha =2$ Eq. (\ref{eq.16bL}) recovers the famous
Wigner's semicircle, $\nu _N^{\left( 2\right) }\left( x\right) =\pi ^{-1}%
\sqrt{2N-x^2}$.

It is worth pointing out that Eqs. (\ref{eq.16aL}) and (\ref{eq.16bL})
obtained within the framework of the theory of orthogonal polynomials
exactly coincide with the density of states calculated for symmetric $\alpha
$-Ensemble in the mean-field approximation \cite{Canalli-Wallin-Kravtsov}.
This circumstance justifies the validity of the mean-field approach for $%
\alpha >1$. For pure linear confinement potential $\left( \alpha =1\right) $
asymptotic formula Eq. (\ref{eq.8L}) fails, signaling that point $\alpha =1$
is a special one.

\section{General properties of symmetric $\alpha $-Ensemble}

The special character of the point $\alpha =1$ may be understood appealing
to the recent mathematical literature on the theory of orthogonal
polynomials with respect to the Freud weights \cite{Levin and Lubinsky}.
Noting that the density of levels $\nu _N$ for random-matrix ensemble with
confinement potential $V\left( x\right) $ is related to an inverse
Christoffel function $\lambda _N^{-1}\left( x\right)
=\sum_{i=0}^{N-1}P_i^2\left( x\right) $ for polynomials $P_i\left( x\right) $
orthogonal with respect to the weight $w\left( x\right) =\exp \left(
-2V\left( x\right) \right) $ as
\begin{equation}
\nu _N\left( x\right) =e^{-2V\left( x\right) }\lambda _N^{-1}\left( x\right)
\text{,}  \label{eq.01a}
\end{equation}
we conclude that the crucial changes occur in the $N$-dependence of the
level density at the origin\cite{Levin and Lubinsky}:
\begin{equation}
\nu _{orig}\left( N,\alpha \right) \sim \left\{
\begin{array}{cc}
N^0, & 0<\alpha <1 \\
\ln N, & \alpha =1 \\
N^{1-1/\alpha }, & \alpha >1
\end{array}
\right. .  \label{eq.01}
\end{equation}
Thus, at the point $\alpha =1$ the functional dependence of the density of
states on the number of levels differs from that both for $0<\alpha <1$ and $%
\alpha >1$. The ratio
\begin{equation}
\xi \left( \alpha \right) =\lim_{N\rightarrow \infty }\frac{\nu _{bg}\left(
N,\alpha \right) }{\nu _{orig}\left( N,\alpha \right) }\text{,}
\label{eq.02}
\end{equation}
with $\nu _{bg}$ being the background density of levels, demonstrates all
the more dramatic behavior. The $\nu _{bg}$ can be estimated as $N/2D\left(
N,\alpha \right) $, where $D\left( N,\alpha \right) $ is the band edge for
symmetric $\alpha $-Ensemble. The estimates of Christoffel functions\cite
{Levin and Lubinsky} allow us to relate $D\left( N,\alpha \right) $ to the
maximal zero $x_{1N}$ of the corresponding $N$-th orthogonal polynomials, $%
x_{1N}\sim N^{1/\alpha }$, so that $\nu _{bg}\sim N^{1-1/\alpha }$. Such a
definition of $\nu _{bg}$ is relevant not only in the case of strong level
confinement, $\alpha >1$, when density of levels in the spectrum bulk is
smooth, but also in the case of weak level confinement although the bulk
level density is no longer constant. In the latter case the density of
levels in the spectrum bulk is very well approximated by $1/\left| x\right|
^{1-\alpha }$ \cite{Canalli-Wallin-Kravtsov}. Since in the bulk of the
spectrum $\left| x\right| =\epsilon D\left( N,\alpha \right) $ with $%
0<\epsilon <1$ we obtain $\nu _{bg}\left( N,\alpha \right) \sim \epsilon
^{\alpha -1}D\left( N,\alpha \right) ^{\alpha -1}\sim N^{1-1/\alpha }$. This
estimates is fully consistent with the definition $\nu _{bg}=N/2D\left(
N,\alpha \right) $ given above. Then we immediately obtain from Eq. (\ref
{eq.01}) that $\xi \left( \alpha \right) =0$ for $0<\alpha \leq 1$. The case
$\alpha >1$ can be treated with the aid of Eq. (\ref{eq.16bL}) which yields
the density of levels at the origin $\left( x=0\right) $
\begin{equation}
\nu _{orig}\left( N,\alpha \right) =\frac{\alpha N^{1-1/\alpha }}{\pi \left(
\alpha -1\right) D\left( \alpha \right) }\text{,}  \label{eq.03b}
\end{equation}
so that $\xi \left( \alpha \right) =\frac \pi 2\left( 1-1/\alpha \right) $
for $\alpha >1$. Finally, we obtain
\begin{equation}
\xi \left( \alpha \right) =\left\{
\begin{array}{c}
0,\text{ }0<\alpha \leq 1 \\
\frac \pi 2\left( 1-\frac 1\alpha \right) ,\text{ }\alpha >1
\end{array}
\right. \text{.}  \label{eq.03}
\end{equation}
The function $\xi \left( \alpha \right) $ is plotted in the Fig. 1.

Equation (\ref{eq.03}) implies that the transition point $\alpha =1$
corresponds to the formation of a sharp peak at the spectrum origin which
also holds for $0<\alpha <1$ (soft confinement), but is absent in the case
of strong confinement potential, $\alpha >1$.

\section{{\bf Border level confinement }$\left( \alpha =1\right) $}

\subsection{\bf Pollaczek random-matrix ensemble (PRME)}

The formalism developed in Section II cannot be directly applied to the case
$\alpha =1$, since the relevant pointwise asymptotic formula Eq. (\ref{eq.8L}%
) fails. Nevertheless, this difficulty can be avoided by choosing $V\left(
x\right) $ in the form
\begin{equation}
V^{\left( \lambda \right) }\left( x\right) =-\frac 12\ln w^{\left( \lambda
\right) }\left( x\right) =\frac 12\sum_{k=0}^\infty \ln \left( 1+\frac{x^2}{%
\left( k+\lambda \right) ^2}\right) +V^{\left( \lambda \right) }\left(
0\right) ,  \label{eq.8}
\end{equation}
where $w^{\left( \lambda \right) }\left( x\right) $ is the weight function
for Pollaczek polynomials. Their basic properties are collected in Appendix
A.

The behavior of confinement potential $V^{\left( \lambda \right) }\left(
x\right) $ can easily be obtained from Eqs. (\ref{eq.8}) and (A2). In the
vicinity of the origin $x=0$, $\left| x\right| \ll \lambda $, expansion of
the Eq. (\ref{eq.8}) yields the quadratic potential
\begin{equation}
V^{\left( \lambda \right) }\left( x\right) \approx V^{\left( \lambda \right)
}\left( 0\right) +\frac 12x^2\Psi ^{\left( 1\right) }\left( \lambda \right) ,
\label{eq.9}
\end{equation}
where $\Psi ^{\left( 1\right) }\left( \lambda \right) =\sum_{k=0}^\infty
\left( k+\lambda \right) ^{-2}$ is a trigamma function. The long-range
behavior of $V^{\left( \lambda \right) }\left( x\right) $ follows from the
asymptotic representation of $\left| \Gamma \left( \lambda +ix\right)
\right| $ when $\left| x\right| \rightarrow \infty $ (see Eq. (A2)), and
turns out to be
\begin{equation}
V^{\left( \lambda \right) }\left( x\right) \approx \frac \pi 2\left|
x\right| -\left( \lambda -\frac 12\right) \ln \left| x\right| -V_\infty
^{\left( \lambda \right) }  \label{eq.10}
\end{equation}
with $V_\infty ^{\left( \lambda \right) }=\lambda \ln 2.$

Thus, the confinement potential in PRME exhibits such a long-range behavior
which involves, in particular, the precisely linear growth at large $\left|
x\right| $ if one puts $\lambda =1/2$. The calculations show (see Fig. 2)
that only small discrepancy between linear confinement potential $V_L\left(
x\right) =\pi \left| x\right| /2$ and $V^{\left( 1/2\right) }\left( x\right)
$ takes place in the region $\left| x\right| <1$, which is negligible as
compared with the band edge $D\left( N,1\right) =N$. This circumstance
allows us to correlate PRME with the transition point $\alpha =1$ of the $%
\alpha $-Ensemble mentioned in the Introduction. The advantage of the
proposed new unitary ensemble of random matrices is that it can be treated
{\it exactly} in the terms of orthogonal Pollaczek polynomials. [For the
sake of convenience, the derivation of asymptotic properties of these
polynomials is entered in Appendix B].

\subsection{\bf Density of states}

The density of states can be calculated making use of the asymptotics of
Pollaczek polynomials found in Appendix B. Namely, the density of states for
eigenvalues of PRME reads \cite{Mehta}
\begin{equation}
\nu _N^{\left( \lambda \right) }\left( x\right) =e^{-2V^{\left( \lambda
\right) }\left( x\right) }\frac{k_{N-1}}{k_Nh_{N-1}^{\left( \lambda \right) }%
}\left( P_{N-1}^{\left( \lambda \right) }\left( x\right) \frac d{dx}%
P_N^{\left( \lambda \right) }\left( x\right) -P_N^{\left( \lambda \right)
}\left( x\right) \frac d{dx}P_{N-1}^{\left( \lambda \right) }\left( x\right)
\right) \text{.}  \label{eq.25}
\end{equation}
Bearing in mind Eqs. (A2), (A4), (A5) and different asymptotics for $%
P_n^{\left( \lambda \right) }\left( x\right) $ near the origin of the
spectrum (Eq. (B14)) and in its bulk (Eq. (B9)), we easily obtain for finite
$\lambda $ in the limit $N\gg 1$:
\begin{equation}
\nu _N^{\left( \lambda \right) }\left( x\right) =\frac 1\pi \left( \ln
\left( 2N\right) -Re\Psi \left( \lambda +ix\right) \right) \text{, }\left|
x\right| \ll \sqrt{2N}\text{,}  \label{eq.26a}
\end{equation}
\begin{equation}
\nu _N^{\left( \lambda \right) }\left( x\right) =\frac 1{2\pi }\ln \left(
\frac{1+\sqrt{1-\left( x/N\right) ^2}}{1-\sqrt{1-\left( x/N\right) ^2}}%
\right) \text{, }1\ll \left| x\right| <N\text{.}  \label{eq.26b}
\end{equation}
Here $\Psi \left( z\right) =\left( d/dz\right) \ln \Gamma \left( z\right) $
is a digamma function.

Equations (\ref{eq.26a}) and (\ref{eq.26b}) lead to the conclusion that the
density of states for PRME does not depend on the parameter $\lambda $ in
the bulk of spectrum, whereas at the origin this $\lambda $-dependence
holds. Such a behavior of the density of states is not a surprise and known
for the generalized Gaussian and Laguerre ensembles \cite{Nagao}.

As $N\rightarrow \infty $ the density of states at the origin tends
asymptotically to the value $\nu _N^{\left( \lambda \right) }\left( 0\right)
=(\ln \left( 2N\right) -\Psi \left( \lambda \right) )/\pi $. We note that
this result is in agreement with $\nu _{orig}\left( N,1\right) $, obtained
for strictly linear potential $V_L\left( x\right) $, see Eq. (\ref{eq.01}).
The non-asymptotic formula for $\nu _N^{\left( \lambda \right) }\left(
0\right) $ that is

\begin{equation}
\nu _N^{\left( \lambda \right) }\left( 0\right) =e^{-2V^{\left( \lambda
\right) }\left( 0\right) }\sum_{j=0}^{N-1}\frac 1{h_j^{\left( \lambda
\right) }}\left| P_j^{\left( \lambda \right) }\left( 0\right) \right| ^2
\label{eq.28}
\end{equation}
can also be obtained. From recurrence Eq. (A1) it follows that

\begin{equation}
P_{2j+1}^{\left( \lambda \right) }\left( 0\right) =0\text{, }P_{2j}^{\left(
\lambda \right) }\left( 0\right) =\left( -1\right) ^j\frac{\Gamma \left(
j+\lambda \right) }{\Gamma \left( \lambda \right) \Gamma \left( j+1\right) }%
\text{.}  \label{eq.29}
\end{equation}
Therefore we arrive at the relationship

\begin{equation}
\nu _N^{\left( \lambda \right) }\left( 0\right) =\frac 1\pi
\sum_{j=0}^{\left[ \frac{N-1}2\right] }\frac{\Gamma \left( j+\frac 12\right)
\Gamma \left( j+\lambda \right) }{\Gamma \left( j+1\right) \Gamma \left(
j+\lambda +\frac 12\right) }\text{.}  \label{eq.30}
\end{equation}
Here Eqs. (A2) and (A4) were used, and $\left[ m\right] $ stands for integer
part of $m$.

Note that the density of states given by Eq. (\ref{eq.26b}), being extended
onto the whole interval $x\in \left[ -N,N\right] $ of the eigenvalues of
PRME, has logarithmic singularity at the origin, but it still remains
integrable and obeys normalization condition
\begin{equation}
\int_{-N}^{+N}dx\nu _N^{\left( \lambda \right) }\left( x\right) =N\text{.}
\label{eq.27}
\end{equation}
Moreover, Eq. (\ref{eq.26b}) exactly coincides with the density of states
for confinement potential $V_L\left( x\right) =\pi \left| x\right| /2$ which
can be found within the mean-field approximation \cite
{Canalli-Wallin-Kravtsov}. This circumstance is a strong evidence that the
mean-field approach, which was proved to be valid for $V\left( x\right) \sim
\left| x\right| ^\alpha $ with $\alpha >1$ (see Section II), is justified
for weaker {\it symmetric} potentials up to linear-like (except for the
region closed to the origin of the spectrum).

The density of states is presented in the Fig. 3, where an excellent
coincidence is observed between analytical asymptotic expressions Eqs. (\ref
{eq.26a}), (\ref{eq.26b}) and the density of states, calculated from the
Hypergeometric representation of Pollaczek polynomials (see Eqs. (A2) and
(A6)):
\begin{equation}
\nu _N^{\left( \lambda \right) }\left( x\right) =\frac{2^{2\lambda -1}}\pi
\left| \Gamma \left( \lambda +ix\right) \right| ^2\sum_{n=0}^{N-1}\left| {}{}%
\frac{\left( 2\lambda \right) _n}{n!}\,_2{\bf F}_1(-n,\lambda +ix;2\lambda
;2)\right| ^2  \label{eq.70aa}
\end{equation}

We also computed the density of states for pure linear confinement potential
$V_L\left( x\right) $ using the well-known in the theory of orthogonal
polynomials matrix representation for Christoffel function \cite{Szego},
\[
\widetilde{\nu }_N\left( x\right) =-e^{-2V_L\left( x\right) }\det \left(
\begin{array}{ccccc}
\mu _0 & \mu _1 & ... & \mu _{N-1} & 1 \\
\mu _1 & \mu _2 & ... & \mu _N & x \\
. & . & . & . & . \\
\mu _{N-1} & \mu _N & ... & \mu _{2N-1} & x^{N-1} \\
1 & x & ... & x^{N-1} & 0
\end{array}
\right) \times
\]
\begin{equation}
\left[ \det^{}\left(
\begin{array}{cccc}
\mu _0 & \mu _1 & ... & \mu _{N-1} \\
\mu _1 & \mu _2 & ... & \mu _N \\
. & . & . & . \\
\mu _{N-1} & \mu _N & ... & \mu _{2N-1}
\end{array}
\right) \right] ^{-1}\text{,}  \label{eq.71}
\end{equation}
where
\begin{equation}
\mu _k=\int_{-\infty }^{+\infty }x^k\exp \left( -2V_L\left( x\right) \right)
dx=\frac{\left[ 1+\left( -1\right) ^k\right] }{\pi ^{k+1}}\Gamma \left(
k+1\right) \text{.}  \label{eq.72}
\end{equation}
The results plotted in Fig. 4 constitute additional justifications of the
use of confinement potential $V^{\left( 1/2\right) }\left( x\right) $
instead of initial linear one, $V_L\left( x\right) $.

\subsection{\bf Two-point correlators}

The two-point kernel can be determined as \cite{Mehta}

\begin{equation}
K_N^{\left( \lambda \right) }\left( x,y\right) =e^{-V^{\left( \lambda
\right) }\left( x\right) -V^{\left( \lambda \right) }(y)}\frac{k_{N-1}}{%
k_Nh_{N-1}^{\left( \lambda \right) }}\frac{P_{N-1}^{\left( \lambda \right)
}\left( x\right) P_N^{\left( \lambda \right) }\left( y\right) -P_N^{\left(
\lambda \right) }\left( x\right) P_{N-1}^{\left( \lambda \right) }\left(
y\right) }{y-x}\text{.}  \label{eq.40}
\end{equation}
Here, again, different asymptotics must be used for the bulk of the spectrum
and for its origin.

In the bulk of the spectrum, $1\ll x<N$, and $1\ll y<N$, Eq. (B9) gives
\[
K_N^{\left( \lambda \right) }\left( x,y\right) =\frac 1{\pi \left(
y-x\right) }\left\{ \left( \frac{1-\left( y/N\right) ^2}{1-\left( x/N\right)
^2}\right) ^{1/4}\sin \left( \Phi _N^{\left( \lambda \right) }\left(
x\right) \right) \cos \left( \Phi _N^{\left( \lambda \right) }\left(
y\right) \right) -\right.
\]
\begin{equation}
\left( \frac{1-\left( x/N\right) ^2}{1-\left( y/N\right) ^2}\right)
^{1/4}\sin \left( \Phi _N^{\left( \lambda \right) }\left( y\right) \right)
\cos \left( \Phi _N^{\left( \lambda \right) }\left( x\right) \right) -\frac{%
y-x}N\times  \label{eq.41}
\end{equation}
\[
\ \ \left. \left[ \left( 1-(\frac xN)^2\right) \left( 1-(\frac yN)^2\right)
\right] ^{-1/4}\sin \left( \Phi _N^{\left( \lambda \right) }\left( x\right)
\right) \sin \left( \Phi _N^{\left( \lambda \right) }\left( y\right) \right)
\right\} \text{,}
\]
where
\begin{equation}
\Phi _N^{\left( \lambda \right) }\left( x\right) =\frac \pi 4+(N+\lambda
)\arccos \left( \frac xN\right) -\pi x\nu _N^{\left( \lambda \right) }\left(
x\right) \text{.}  \label{eq.42}
\end{equation}

It can be seen that despite of this function has a rather complicated form,
the {\it local} properties of the two-point kernel remain to be universal.
Really, for $\left| x-y\right| $ that is much smaller than the scale $%
\varsigma $ of characteristic changes of the mean level density, $\varsigma
\sim \nu _N^{\left( \lambda \right) }\left( d\nu _N^{\left( \lambda \right)
}/dx\right) ^{-1}\sim N$, and both $x$ and $y$ at finite distance from the
edge of the spectrum, the third term can be neglected. Then, one obtains
\begin{equation}
K_N^{\left( \lambda \right) }\left( x,y\right) =\frac 1{\pi \left(
y-x\right) }\sin \left( \Phi _N^{\left( \lambda \right) }\left( x\right)
-\Phi _N^{\left( \lambda \right) }\left( y\right) \right) \text{.}
\label{eq.43}
\end{equation}
Making use of the different representation for $\Phi _N^{\left( \lambda
\right) }$,
\begin{equation}
\Phi _N^{\left( \lambda \right) }\left( x\right) =\frac \pi 4\left(
1+2N\right) +\lambda \arccos \left( \frac xN\right) -\pi \int_0^x\nu
_N^{\left( \lambda \right) }\left( z\right) dz\text{,}  \label{eq.44}
\end{equation}
with $\nu _N^{\left( \lambda \right) }$ defined by Eq. (\ref{eq.26b}), we
easily obtain the universal form of the two-point kernel in the limit $%
N\rightarrow \infty $:
\begin{equation}
K_N^{\left( \lambda \right) }\left( x,y\right) =\frac 1{\pi \left(
x-y\right) }\sin \left( \pi \overline{\nu }_N^{\left( \lambda \right)
}(x-y)\right) \text{,}  \label{eq.45}
\end{equation}
where $\overline{\nu }_N^{\left( \lambda \right) }=\nu _N^{\left( \lambda
\right) }\left( \left( x+y\right) /2\right) $ is a local mean level density.
Equation (\ref{eq.45}) proves the local universality of the two-point kernel
and is valid on the scale $\left| x-y\right| $ which is much larger than the
mean level-spacing $\left( \overline{\nu }_N^{\left( \lambda \right)
}\right) ^{-1}\sim 1$.

In the vicinity of the spectrum center, $\left| x\right| \ll \sqrt{2N}$, and
$\left| y\right| \ll \sqrt{2N}$, the asymptotic formula Eq. (B14) and Eq. (%
\ref{eq.40}) yield
\begin{equation}
K_N^{\left( \lambda \right) }\left( x,y\right) =\frac 1{\pi \left(
y-x\right) }\sin \left( \phi _N^{\left( \lambda \right) }\left( y\right)
-\phi _N^{\left( \lambda \right) }\left( x\right) \right) \text{,}
\label{eq.46}
\end{equation}
\begin{equation}
\phi _N^{\left( \lambda \right) }\left( x\right) =x\ln \left( 2N\right)
-\arg \Gamma \left( \lambda +ix\right) =\pi \int_0^x\nu _N^{\left( \lambda
\right) }\left( z\right) dz\text{,}  \label{eq.47}
\end{equation}
where $\nu _N^{\left( \lambda \right) }$ is determined by Eq. (\ref{eq.26a}%
). Since near the origin the scale $\varsigma $ of characteristic changes of
the mean level density is of the order $\lambda $, Eq. (\ref{eq.46}) can
also be rewritten in the universal form Eq. (\ref{eq.45}) but with density
of states $\nu _N^{\left( \lambda \right) }$ corresponding to the center of
the spectrum. Here, again, universal form of the two-point kernel is valid
on the scale $\left| x-y\right| $ which is much larger than the mean
level-spacing $\left( \overline{\nu }_N^{\left( \lambda \right) }\right)
^{-1}\sim 1/\ln N$.

Thus, we arrive at the conclusion that two-level cluster function for PRME,
\begin{equation}
Y_2\left( s,s^{\prime }\right) =\left( \frac{K_N^{\left( \lambda \right)
2}\left( x,y\right) }{\nu _N^{\left( \lambda \right) }\left( x\right) \nu
_N^{\left( \lambda \right) }\left( y\right) }\right) _{%
%TCIMACRO{\QATOPD. . {x=x\left( s\right) }{y=y\left( s^{\prime }\right) } }
%BeginExpansion
{x=x\left( s\right)  \atopwithdelims.. y=y\left( s^{\prime }\right) }
%EndExpansion
}\text{,}  \label{eq.61}
\end{equation}
being rewritten in the terms of the eigenvalues measured in the local level
spacing $s=x\overline{\nu }_N^{\left( \lambda \right) }$ and $s^{\prime }=y%
\overline{\nu }_N^{\left( \lambda \right) }$, locally takes the universal
form Eq. (\ref{eq.18L}) for any finite $\lambda $.

The connected correlations between the density of eigenvalues are determined
as
\begin{equation}
\left[ \nu _N^{\left( \lambda \right) }\left( x,y\right) \right]
_{con}=-K_N^{\left( \lambda \right) 2}\left( x,y\right)  \label{eq.62aa}
\end{equation}
if $x\neq y$ and oscillate rapidly on the scale of the band width $D\left(
N,1\right) =N$. The smoothed correlation function can easily be determined.
Bearing in mind Eq. (\ref{eq.41}) we obtain in the bulk of the spectrum
after averaging over rapid oscillations
\begin{equation}
\left[ \nu _N^{\left( \lambda \right) }\left( x,y\right) \right] _{con}=-%
\frac 1{2\pi ^2\left( x-y\right) ^2}\frac{D\left( N,1\right) ^2-xy}{\sqrt{%
\left( D\left( N,1\right) ^2-x^2\right) \left( D\left( N,1\right)
^2-y^2\right) }}\text{, }x\neq y.  \label{eq.62ab}
\end{equation}
The smoothed correlations near the spectrum origin can be calculated by
means of Eqs. (\ref{eq.46}) and (\ref{eq.47}), and turn out to be
\begin{equation}
\left[ \nu _N^{\left( \lambda \right) }\left( x,y\right) \right] _{con}=-%
\frac 1{2\pi ^2\left( x-y\right) ^2}\text{, }x\neq y.  \label{eq.62ac}
\end{equation}
Equation (\ref{eq.62ac}) is the limited case of the Eq. (\ref{eq.62ab}) when
both $\left| x\right| $ and $\left| y\right| $ are much smaller than the
band edge $D\left( N,1\right) $.

Thus, Eqs. (\ref{eq.62ab}) and (\ref{eq.62ac}) prove that smoothed
correlations in PRME follow universal form.

\section{Conclusion}

We have considered the general properties of symmetric $\alpha $-Ensemble
with confinement potential $V\left( x\right) \sim \left| x\right| ^\alpha $
and established that the phenomenon of the peak formation in the density of
states takes place. Namely, we have demonstrated that this sharp peak occurs
for $0<\alpha \leq 1$ and is absent in the opposite case $\alpha >1$. In
this sense the point $\alpha =1$ is a transition point associated with
border level confinement. It has been shown that this transition point may
be explored by means of slight changes in strictly linear confinement
potential $V_L\left( x\right) $ near the spectrum origin leading to the
potential $V^{\left( 1/2\right) }\left( x\right) =-\frac 12\ln w^{\left(
1/2\right) }\left( x\right) =\frac 12\ln \cosh \left( \pi x\right) $ which
is a special case of more general confinement potential Eq. (\ref{eq.8})
connected to the symmetric Pollaczek polynomials.

We have calculated the density of states, two-point kernel, two-level
cluster function and smoothed correlations of the density of eigenvalues in
the large-$N$ limit for $\alpha $-Ensemble with strong and border level
confinement. It has been shown that both properly rescaled two-level cluster
function and smoothed correlations of level density take the universal forms
which are typical of random-matrix ensembles with the Wigner-Dyson level
statistics.

We have also demonstrated that the mean-field approximation is valid for
calculation of the level density in symmetric $\alpha $-Ensemble with strong
level confinement, $\alpha >1$. In the case of border level confinement, $%
\alpha =1$, the mean-field approach is proved to be justified in the bulk of
spectrum, failing near its origin.

We would like to stress that treatment presented in this paper is {\it %
rigorous} and does not appeal to commonly used conjectures and approximate
methods.

\begin{center}
{\bf Acknowledgment}
\end{center}

E. K. is grateful to The Ministry of Science and The Arts of Israel for
financial support.

\newpage\

\begin{center}
{\bf Appendix A}. {\bf Definitions and basic properties of symmetric
Pollaczek polynomials}
\end{center}

Symmetric Pollaczek polynomials $P_n^{(\lambda )}(x)$ are determined by the
recurrence equation \cite{Bateman},\cite{Chihara}
\begin{equation}
nP_n^{(\lambda )}(x)-2xP_{n-1}^{(\lambda )}(x)+(n-2+2\lambda
)P_{n-2}^{(\lambda )}(x)=0,\quad n=1,2,3,...  \eqnum{A1}
\end{equation}
with $P_{-1}^{(\lambda )}(x)=0,P_0^{(\lambda )}(x)=1,$ and $\lambda >0$.
These polynomials are orthogonal in the interval $-\infty <x<\infty $ with
the weight function
\begin{equation}
w^{(\lambda )}(x)=\frac{2^{2\lambda -1}}\pi \left| \Gamma (\lambda
+ix)\right| ^2=\frac{2^{2\lambda -1}\left| \Gamma \left( \lambda \right)
\right| ^2}\pi \prod_{k=0}^\infty \left( 1+\frac{x^2}{\left( k+\lambda
\right) ^2}\right) ^{-1},  \eqnum{A2}
\end{equation}
so that
\begin{equation}
\int_{-\infty }^{+\infty }dxP_n^{(\lambda )}(x)P_m^{(\lambda
)}(x)w^{(\lambda )}(x)=\delta _{nm}h_n^{(\lambda )},  \eqnum{A3}
\end{equation}
\begin{equation}
h_n^{(\lambda )}=\frac{\Gamma (n+2\lambda )}{\Gamma (n+1)}.  \eqnum{A4}
\end{equation}

{}From recurrence equation the leading coefficient of $P_n^{(\lambda )}$ can
be found:

\begin{equation}
k_n=\frac{2^n}{\Gamma \left( n+1\right) }.  \eqnum{A5}
\end{equation}

The following Hypergeometric representation holds for Pollaczek polynomials:

\begin{equation}
P_n^{(\lambda )}(x)=\frac{(2\lambda )_n}{n!}{}{}\,_2{\bf F}_1(-n,\lambda
+ix;2\lambda ;2)\exp \left( i\pi n/2\right) {}.  \eqnum{A6}
\end{equation}

Lastly, we present the generating function that reads
\begin{equation}
F(x,w)=\sum_{k=0}^\infty P_k^{(\lambda )}(x)w^k=\left( \frac{1-iw}{1+iw}%
\right) ^{ix}\frac 1{\left( 1+w^2\right) ^\lambda },\text{ }\left| w\right|
<1.  \eqnum{A7}
\end{equation}
\newpage\

\begin{center}
{\bf Appendix B. Asymptotic formulae for Pollaczek polynomials}
\end{center}

\subsubsection{\bf Formula of Plancherel-Rotach type}

We start with Eq. (A7) which being reversed yields
\begin{equation}
P_n^{(\lambda )}(x)=\frac 1{2\pi i}\oint_{\gamma _0}\frac{dw}{w^{n+1}}%
F\left( x,w\right) \text{.}  \eqnum{B1}
\end{equation}
Here the integration is extended over a contour $\gamma _0$ that encloses
the origin $w=0$ and does not intersect the branch cuts $\left[ i,+i\infty
\right[ $ and $\left] -i\infty ,-i\right] $ associated with the
singularities of generating function. Choosing $x=\left( n+1\right) \cos
\theta $ with $\epsilon \leq \theta \leq \pi -\epsilon $, where $\epsilon $
is a fixed positive number smaller than $\pi /2$, we rewrite Eq. (B1) as
\begin{equation}
P_n^{(\lambda )}(x)=\frac 1{2\pi i}\oint_{\gamma _0}\frac{dw}{\left(
1+w^2\right) ^\lambda }\exp \left[ \left( n+1\right) \left( i\cos \theta \ln
\left( \frac{1-iw}{1+iw}\right) -\ln w\right) \right] \text{.}  \eqnum{B2}
\end{equation}
For the calculation of this contour integral in the limit $n\gg 1$ the
method of steepest descent \cite{Szego},\cite{DeBruijn} can be applied. The
saddle-point condition is
\begin{equation}
S^{^{\prime }}\left( w\right) =\frac \partial {\partial w}\left[ i\cos
\theta \ln \left( \frac{1-iw}{1+iw}\right) -\ln w\right] =0\text{,}
\eqnum{B3}
\end{equation}
whence
\begin{equation}
w_{sp}=e^{\pm i\theta }\text{,}  \eqnum{B4}
\end{equation}
and the original contour $\gamma _0$ of integration enclosing the origin
must be deformed to pass through the points $w=e^{\pm i\theta }$ along the
directions of steepest descent and to avoid intersections with the branch
cuts.

The contributions of both saddle points to the leading term of the
asymptotic expansion of the integral (B2) are of the same order. Therefore
\begin{equation}
P_n^{(\lambda )}(x)=\frac 1{2\pi i}\sum_{w_{sp}}f\left( w_{sp}\right) \sqrt{-%
\frac{2\pi }{\left( n+1\right) S^{^{\prime \prime }}\left( w_{sp}\right) }}%
\exp \left[ \left( n+1\right) S\left( w_{sp}\right) \right] \text{,}
\eqnum{B5}
\end{equation}
where

\begin{equation}
S\left( w_{sp}\right) =\frac \pi 2\left| \cos \theta \right| \mp i\left(
\cos \theta \ln \left| \tan \left( \frac \theta 2-\frac \pi 4\right) \right|
+\theta \right) \text{,}  \eqnum{B6}
\end{equation}
\begin{equation}
S^{^{\prime \prime }}\left( w_{sp}\right) =e^{\mp i\left( 2\theta +\frac \pi
2\right) }\tan \theta {}\text{,}\,  \eqnum{B7}
\end{equation}
and
\begin{equation}
f\left( w_{sp}\right) =\frac 1{\left( 1+w_{sp}^2\right) ^\lambda }=\frac{%
e^{\mp i\theta \lambda }}{\left( 2\cos \theta \right) ^\lambda }\text{.}
\eqnum{B8}
\end{equation}

Making use of Eqs. (B5) - (B8) we obtain the following asymptotic formula:
\[
P_n^{(\lambda )}(x)\approx \frac{\left( n+1\right) ^{\lambda -1}\exp \left(
\pi \left| x\right| /2\right) }{\left( 2\left| x\right| \right) ^{\lambda -%
\frac 12}\sqrt{\pi }{}\left[ 1-\left( x/(n+1)\right) ^2\right] ^{1/4}}\times
\]
\begin{equation}
\sin \left( \frac \pi 4+\left( n+\lambda \right) \arccos \left( \frac x{n+1}%
\right) +\frac x2\ln \left( \frac{1-\sqrt{1-\left( x/(n+1)\right) ^2}}{1+%
\sqrt{1-\left( x/(n+1)\right) ^2}}\right) \right) \text{.}  \eqnum{B9}
\end{equation}

\subsubsection{{\bf Vicinity of the origin: }$\left| x\right| \ll \protect
\sqrt{2n}$}

At the origin $x\sim 0$ the saddle-point approximation used in the previous
section to calculate asymptotic value of the integral (B1) does not work,
and the Plancherel-Rotach type formula fails. In this interval the Darboux
method \cite{Szego} turns out to be fruitful.

To obtain some asymptotic expansion for a polynomial $P_n$ in accordance
with the Darboux Theorem we have to expand the corresponding generating
function $F\left( x,w\right) $ in the vicinities of its singularities $%
e^{i\phi _k}$ on the unit circle $\left| w\right| =1$ into series of the
form
\begin{equation}
F\left( x,w\right) =\sum_{m=0}^\infty c_m^{\left( k\right) }\left(
1-we^{-i\phi _k}\right) ^{a_k+mb_k}\text{.}  \eqnum{B10}
\end{equation}
Then the expression
\begin{equation}
\sum_{m=0}^\infty \sum_kc_m^{\left( k\right) }%
%TCIMACRO{\binom{a_k+mb_k}n }
%BeginExpansion
{a_k+mb_k \choose n}
%EndExpansion
\left( -e^{i\phi _k}\right) ^n  \eqnum{B11}
\end{equation}
furnishes an asymptotic expansion for $P_n(x)$.

For Pollaczek polynomials the singularities of the generating function Eq.
(A7) occur at the points $w=\pm i$, in whose vicinities generating function
can be expanded as
\begin{equation}
F\left( x,w\right) =2^{-\lambda \pm ix}\sum_{m=0}^\infty
%TCIMACRO{\binom{-\lambda \pm ix}m }
%BeginExpansion
{-\lambda \pm ix \choose m}
%EndExpansion
\left( -\frac 12\right) ^m\left( 1\pm iw\right) ^{m-\lambda \pm ix}\text{.}
\eqnum{B12}
\end{equation}
Then the expression
\begin{equation}
\left( -1\right) ^nRe~2^{ix-\lambda +1}\sum_{m=0}^\infty
%TCIMACRO{\binom{ix-\lambda }m }
%BeginExpansion
{ix-\lambda  \choose m}
%EndExpansion
%TCIMACRO{\binom{m-\lambda -ix}n }
%BeginExpansion
{m-\lambda -ix \choose n}
%EndExpansion
\frac{e^{i\frac \pi 2\left( 2m-n\right) }}{2^m}  \eqnum{B13}
\end{equation}
leads to the asymptotic formula for Pollaczek polynomials.

The leading term results from $m=0$ in Eq. (B13), so that for finite $%
\lambda $ and $n\gg 1$%
\begin{equation}
P_n^{(\lambda )}(x)\approx \left( \frac n2\right) ^{\lambda -1}\frac 1{%
\left| \Gamma \left( \lambda +ix\right) \right| }\cos \left( x\ln \left(
2n\right) -\arg \Gamma \left( \lambda +ix\right) -\frac \pi 2n\right) \text{.%
}  \eqnum{B14}
\end{equation}
The next term $\left( m=1\right) $ of asymptotic expansion Eq. (B13) is of
the order ${\cal O}\left( \sqrt{\left( x^2+\lambda ^2\right) \left(
x^2+\left( \lambda -1\right) ^2\right) }/2n\right) $. This circumstance
imposes the restrictions $\left| \lambda \right| \ll \sqrt{2n}$ and $\left|
x\right| \ll \sqrt{2n}$ for Eq. (B14).

\newpage\

\begin{center}
{\bf Figure captions}
\end{center}

Fig. 1. Function $\xi \left( \alpha \right) $ demonstrating the sharp peak
formation in the density of states at the origin of the spectrum when $%
\alpha \leq 1$

Fig. 2. Confinement potential $V^{\left( 1/2\right) }\left( x\right)
+V_\infty ^{\left( 1/2\right) }$ for Pollaczek random-matrix ensemble with $%
\lambda =1/2$ (dashed line) and linear potential $V_L\left( x\right) =\pi
\left| x\right| /2$ (solid line). The only small discrepancy takes place in
the narrow region $\left| x\right| <1$

Fig. 3. Density of levels for PRME. Solid line: asymptotic formulae Eqs.
(30) and (31). Dotted line: Hypergeometric representation Eq. (36).
Parameters: $\lambda =1/2$, $N=20$

Fig. 4. Density of levels for confinement potential $V_L\left( x\right) =\pi
\left| x\right| /2$ calculated from Eqs. (37) and (38) (dotted line) and for
Pollaczek random-matrix ensemble with $\lambda =1/2$ (Eqs. (30) and (31),
solid line). $N=20$

\newpage\

$^{\#}$On leave from: Institute for Low Temperature Physics and Engineering,
Kharkov 310164, Ukraine


\begin{references}
\bibitem{Mehta}  M. L. Mehta, {\it Random matrices}, (Academic Press,
Boston, 1991)

\bibitem{Shklovskii and Kravtsov}  B. I. Shklovskii {\it et al, Phys. Rev. B.%
}, {\bf 47}, 11487 (1993); V. E. Kravtsov, I. V. Lerner, B. L. Altshuler,
and A. G. Aronov, {\it Phys. Rev. Lett.}, {\bf 72}, 888 (1994)

\bibitem{Shapiro}  J.-L. Pichard, B. Shapiro, {\it J. Phys.},{\bf \ 4}, 623
(1994); M. Moshe, H. Neuberger, and B. Shapiro, {\it Phys. Rev. Lett.}, {\bf %
73}, 1497 (1994); M. Kreynin, B. Shapiro, {\it Phys. Rev. Lett.}, {\bf 74},
4122 (1995)

\bibitem{Muttalib}  K. A. Muttalib, Y. Chen, M. E. H. Ismail, and V. N.
Nicopoulus, {\it Phys. Rev. Lett.}, {\bf 71}, 471 (1993); Y. Chen, M. E. H.
Ismail, and K. A. Muttalib, {\it J. Phys. Cond. Matter}, {\bf 5}, 177 (1993)

\bibitem{Canalli-Kravtsov's reprint}  C. M. Canalli, V. E. Kravtsov, {\it %
Phys. Rev. E., }{\bf 51}, R5185 (1995)

\bibitem{Manning}  Y. Chen, S. M. Manning, {\it J. Phys. Cond. Matter}, {\bf %
6}, 3039 (1994); see also Ref. \cite{Canalli-Wallin-Kravtsov}

\bibitem{Canalli-Wallin-Kravtsov}  C. M. Canalli, M. Wallin, V. E. Kravtsov,%
{\it \ Phys. Rev. B.}, {\bf 51}, 2831 (1995)

\bibitem{FKY}  V. Freilikher, E. Kanzieper, and I. Yurkevich, to be published

\bibitem{Brezin-Zee}  E. Br\'ezin, A. Zee, {\it Nucl. Physics B}[FS], {\bf %
402}, 613 (1993)

\bibitem{Beenakker}  C. W. J. Beenakker, {\it Nucl. Physics B}[FS], {\bf 422}%
, 515 (1994)

\bibitem{Weidenmuller}  G. Hackenbroich, H. A. Weidenm\"uller, {\it Phys.
Rev. Lett.}, {\bf 74}, 4118 (1995)

\bibitem{Nevai}  P. Nevai, {\it J. Appr. Theory}, {\bf 48}, 3 (1986)

\bibitem{Lubinsky}  D. S. Lubinsky, {\it Acta Appl. Math.}, {\bf 33}, 121
(1993)

\bibitem{Rahmanov}  E. A. Rahmanov, {\it Strong Asymptotics for Orthogonal
Polynomials with Exponential Weights on }{\bf R}, in: {\it Springer Lecture
Notes in Mathematics}, vol. 1550, p. 71 (Eds. A. A. Gonchar and E. B. Saff,
Nauka, Moscow, 1992)

\bibitem{Levin and Lubinsky}  A. L. Levin, D. S. Lubinsky, {\it Constr.
Approx., }{\bf 8}, 463 (1992); {\it J. Approx. Theory,}{\bf \ 80}, 219 (1995)

\bibitem{Nagao}  T. Nagao, K. Slevin, {\it J. Math. Phys.}, {\bf 34}, 2075
(1993)

\bibitem{Szego}  G. Szeg\"o, {\it Orthogonal polynomials}, (American
Mathematical Society, Providence, 1967), vol. XXIII, Ch. 8

\bibitem{Bateman}  Bateman Manuscript Project (Director: A. Erd\'elyi), {\it %
Higher transcendental functions}, (McGraw-Hill, New York-Toronto-London,
1953), vol. 2, Ch. 10

\bibitem{Chihara}  T. S. Chihara, {\it An introduction to orthogonal
polynomials}, (Gordon and Breach Science Publishers, New York-London-Paris,
1978), Ch. 6

\bibitem{DeBruijn}  N. G. De Bruijn, {\it Asymptotic methods in analysis},
(North-Holland Publishing Co., Amsterdam, 1958)
\end{references}
\end{document}